\documentclass[prl,twocolumn,showpacs,superscriptaddress,amsmath,amssymb]{revtex4}

\usepackage{graphicx}%
\usepackage{dcolumn}%

\begin{document}

\title{What the resonance peak does.}

\author{S. V. Borisenko}
\affiliation{Institute for Solid State Research, IFW-Dresden, P.O.Box 270116, D-01171 Dresden, Germany}

\author{A. A. Kordyuk}
\affiliation{Institute for Solid State Research, IFW-Dresden, P.O.Box 270116, D-01171 Dresden, Germany}
\affiliation{Institute of Metal Physics of National Academy of Sciences of Ukraine, 03142 Kyiv, Ukraine}

\author{T. K. Kim}
\author{A. Koitzsch}
\author{M. Knupfer}
\author{M. S. Golden}
\altaffiliation[New address: ]{Van der Waals-Zeeman Institute, University of Amsterdam, The Netherlands}
\author{J. Fink}
\affiliation{Institute for Solid State Research, IFW-Dresden, P.O.Box 270116, D-01171 Dresden, Germany}

\author{M. Eschrig}
\affiliation{Institut f\"ur Theoretische Festk\"orperphysik, Universit\"at Karlsruhe, D-76128 Karlsruhe, Germany}

\author{H. Berger}
\affiliation{Institute of Physics of Complex Matter, EPFL, CH-1015 Lausanne, Switzerland}

\author{R. Follath}
\affiliation{BESSY GmbH, Albert-Einstein-Strasse 15, 12489 Berlin, Germany}

\date{\today}
\begin{abstract}
High-resolution angle-resolved photoemission with variable excitation energies 
is used to disentangle bilayer splitting effects and intrinsic (self-energy) 
effects in the electronic spectral function near the ($\pi$,0)-point of differently doped 
(Pb,Bi)$_2$Sr$_2$CaCu$_2$O$_{8+\delta}$. 
In contrast to overdoped samples, where intrinsic effects at the ($\pi$,0)
point are virtually absent, we find in underdoped samples \textit{intrinsic} effects
in the superconducting-state ($\pi$,0) spectra of the antibonding band.
This intrinsic effect is present only below the critical temperature and
weakens considerably with doping. Our results give strong support for
models which involve a strong coupling of electronic excitations with
the resonance mode seen in inelastic neutron scattering experiments.
\end{abstract}

\maketitle
The sharp magnetic resonance peak observed in inelastic neutron scattering (INS) experiments \cite{Rossat,Tranquada,Fong,Dai}, is unanimously considered to be one of the most striking features of the high temperature superconducting cuprates which suggests an important role of magnetism in the mechanism of the HTSC \cite{Kee,Abanov}. It has been argued 
\cite{Norman,Abanov} that the emergence of the resonance below the critical temperatures ($T_c$) has a strong feedback effect on the electronic properties of the cuprates, leading to the appearance of the peak-dip-hump (PDH) features in the angle-resolved photoemission (ARPES) spectra near ($\pi$, 0)-points \cite{Dessau}, kinks in the dispersion along the nodal direction \cite{BKJ} and a dip at characteristic energies in the SIS tunneling conductance \cite{Zasad} or in the optical conductivity \cite{Basov}.
Such spectroscopic evidence for the sensitivity of the charge dynamics to the spin-excitations below $T_c$ strongly supports a model
of magnetically-mediated superconductivity based on spin-fluctuation exchange \cite{Chubukov}. 
Indeed, there are even proposals which consider the resonance mode as a boson which mediates the pairing itself \cite{Carbotte}. 
From another point of view, however, the resonance is only a measure of pairing and phase coherence \cite{Dai} and due to its small spectral weight is not 
able to be the "glue" in any conventional pairing theory 
\cite{Abanov}. In extreme case the resonance is argued not even beeing able to
account for the anomalies observed in ARPES and optical absorption data \cite{Kee}.

To complicate matters further, recent ARPES data appear to weight in on both sides of the debate. On the one hand, it has been shown that for overdoped HTSC the famous PDH line shape of the superconducting-state ($\pi$, 0) ARPES spectrum cannot be taken as a signature of the coupling to the resonant mode, but is rather due to the superposition of two bilayer-split bonding and antibonding bands \cite{KordPRL}. On the other hand, the renormalisation of the electronic dispersion near the "antinodal" points of the normal state Fermi surface (located some $18\%$ of the ($\pi$, 0)-($\pi$, $\pi$) distance away from the ($\pi$, 0)-point) in the superconducting state in overdoped samples has been suggested to be a sign of strong coupling of the electronic system to the magnetic resonance mode \cite{Gromko1,Gromko2}. 
The former observation does not contradict the latter \cite{Eschrig_Norman}
since at the ($\pi$, 0)-point in the overdoped case the antibonding band is too close to the Fermi level to be strongly influenced by the mode. Such situation naturally focuses one's attention on the underdoped compounds where the saddle point of the antibonding band is known to be at higher binding energies \cite{Kord_gap}. Considering that it is the ($\pi$, 0)-point where the electron density 
of states
is maximal and equivalent points are separated by a ($\pi$, $\pi$)-vector, the coupling to the magnetic resonance mode is expected to be the strongest there and it is imperative that feedback effects on the spectral function of underdoped systems are examined in detail in this region. If a similar picture regarding the 
origin of the PDH line shape of the ($\pi$, 0)-spectrum 
as in overdoped compounds is observed in underdoped 
compounds, then the concept of spin-mediated pairing looses one of its 
strongest supporting arguments coming from the experiment.

In this Letter we show that the situation in the underdoped regime is different.
While spectra using low (19--22.4 eV) photon energies are complicated by
a superposition of bilayer splitting effects and possible intrinsic effects,
the spectra with virtually no contribution from the bonding band (e.g. $h\nu$ = 29 or 50 eV), i.e. representing purely antibonding component, demonstrate a clear evidence for an \textit{intrinsic} anomaly which cannot be accounted 
for by a simple spectral function but could be well explained by taking into account self energy effects originating from the coupling of electrons to a sharp collective mode.
Furthermore, extracted from our experimental data characteristics of the mode, 
such as its momentum, temperature and doping dependencies as well as energy 
match the characteristics of the resonance peak observed in INS.

The ARPES experiments were carried out using radiation from U125/1-PGM beam line and angle-multiplexing photoemission spectrometer (SCIENTA SES100) at BESSY synchrotron radiation facility. The total energy resolution ranged from 8 meV (FWHM) at $h\nu=$ 17--25 eV to 22.5 meV at $h\nu=$ 65 eV. Data were collected on under- ($T_c$=77K) and overdoped ($T_c$=69K) single crystals of Pb-Bi2212 \cite{Footnote_PbDoping}. All ($\pi$,0) energy distribution curves (EDCs), unless other is specified in the text, were measured at a temperature of 30K - deep in the superconducting state.

\begin{figure}[t!]
\includegraphics[width=8.47cm]{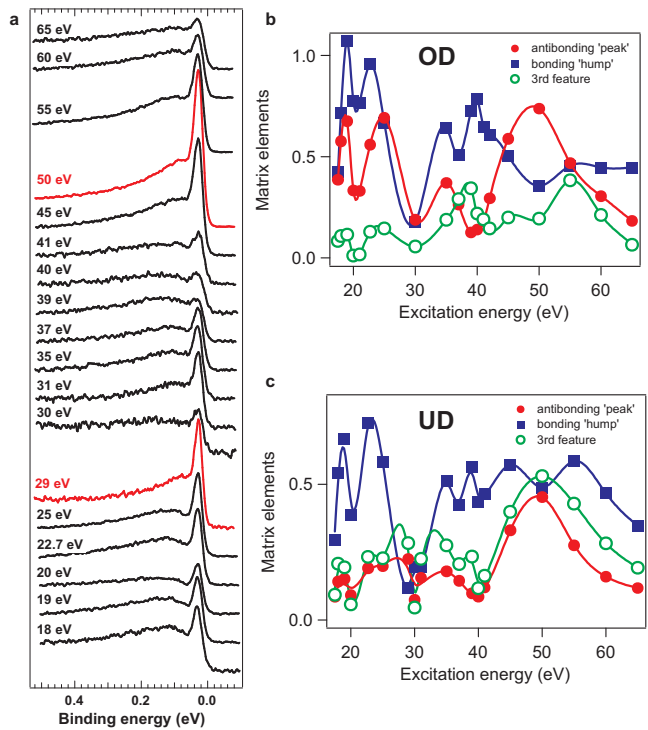}%
\caption{\label{Spectra} a) The ($\pi$, 0) photoemission spectra from the superconducting state of underdoped ($T_c$=77 K) sample for different excitation energies. Panels (b) and (c) show the results of the fitting procedure described in Ref.\onlinecite{KordPRL}, giving the intensity prefactors M$_a$, M$_b$, and M$_c$ as functions of the excitation energy for overdoped (Ref.\onlinecite{KordPRL}) and underdoped Pb-Bi2212, respectively.} 
\end{figure}

We begin with presenting ($\pi$, 0)-spectra measured using different excitation energies in Fig.~\ref{Spectra}(a). At first glance, a comparison of these experimental data with analogous data from overdoped Pb-Bi2212 in Ref.\onlinecite{KordPRL} immediately suggests a similar scenario - the PDH line shape is strongly excitation energy dependent and therefore cannot be considered as originating from a single 
spectral function. One easily notes the varying relative intensity of the low-energy (peak) and the high-energy (hump) features which could be naturally assumed to be the consequence of the different emission probability (matrix elements) from the separate bands. Moreover, the excitation energy dependence of the relative intensity qualitatively agrees with the one observed in the overdoped regime, as one can intuitively expect for the split pair of bands of the same atomic character.

However, a closer inspection of Fig.~\ref{Spectra}(a) reveals that there is an important and noticeable difference with respect to the data from overdoped
crystals \cite{KordPRL}. While the overdoped datasets were characterised by the fact that a significant number of spectra exhibited no dip in the lineshape at all, such smooth, 'dip-less' spectra are remarkable by their absence in the underdoped data. Every spectrum in Fig.~\ref{Spectra}(a) possesses either a dip or a plateau feature (e.g. $h\nu=$29 eV and $h\nu=$50 eV spectra) which separates the high and low energy parts of the spectral profile. We now move beyond this qualitative description by fitting the ($\pi$, 0)-spectra with three features (plus a background), as was done for the overdoped case \cite{KordPRL,gap}. In Fig.~\ref{Spectra}(c) we plot the photon energy dependence of the intensity prefactors of each of the three components of the fit, $M_a$, $M_b$ and $M_c$, together with data from overdoped Pb-Bi2212 [Ref.~\onlinecite{KordPRL}] in Fig.~\ref{Spectra}(b).

There is a global agreement between the behavior of the matrix elements of the 'hump' and 'peak' in the underdoped and overdoped samples. This immediately indicates that all arguments issued in Ref.~\onlinecite{KordPRL} regarding the assignment of these features to the bonding and antibonding bands in overdoped regime are fully applicable here: on the "large scale" PDH line shape is due to the superposition of these two components. What is really different between Figs.~\ref{Spectra}(b) and \ref{Spectra}(c) is the behavior of the third feature of the fit (M$_c$). While in the upper panel M$_c$ is relatively small and its energy dependence barely tracks that of either of the other two features, in the lower panel we see the striking similarity between $M_a$ and $M_c$, i.e. between the peak and the third feature. Such a close similarity implies that these two features are components of the same, single spectral feature which possesses a more complex lineshape. Moreover, it is easy to see from the Fig.~\ref{Spectra}(c) that considering the third feature and the peak as constituents of the antibonding spectral function one gains better quantitative agreement between their total spectral weight and the spectral weight of the antibonding band in the overdoped case (see Fig.~\ref{Spectra}(b)).

\begin{figure}[t!]
\includegraphics[width=8.4cm]{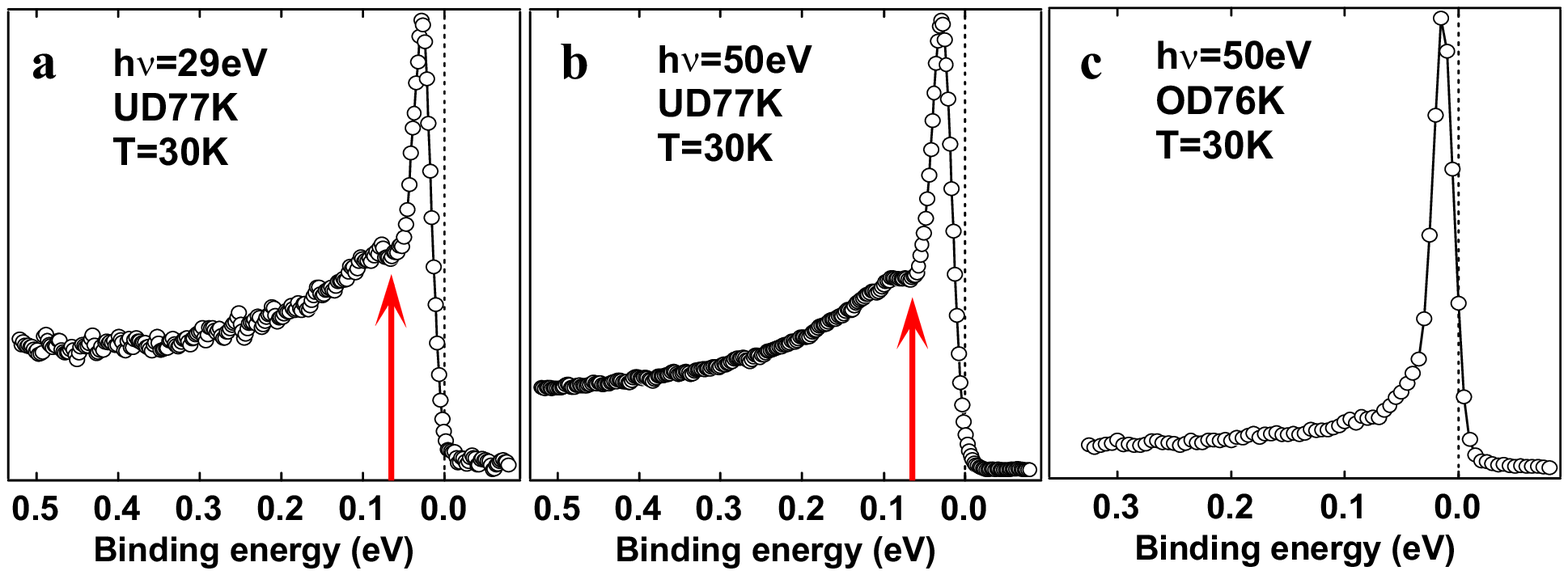}\\%
\includegraphics[width=8.5cm]{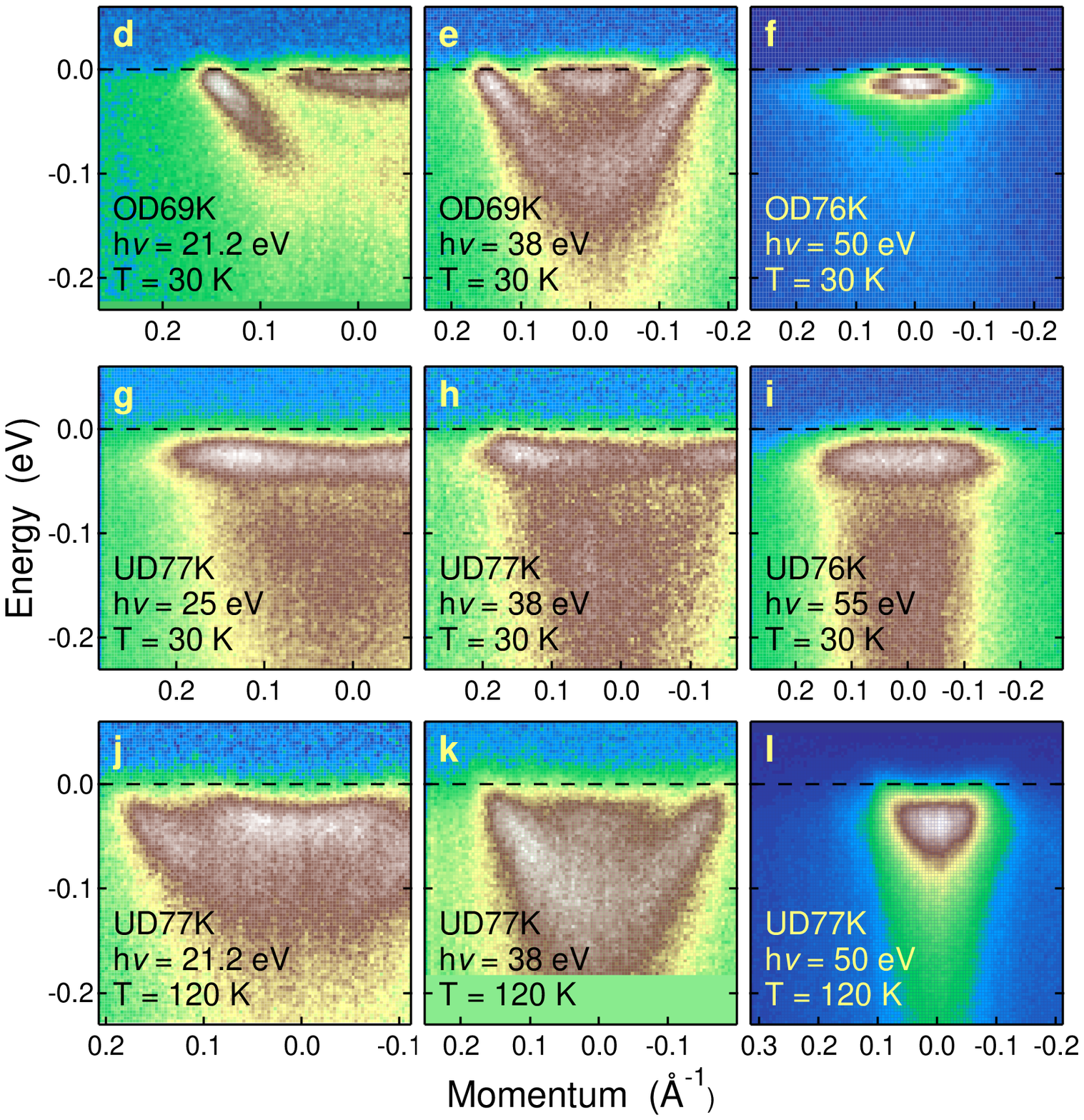}%
\caption{\label{EDCs&EDMs} (a)-(c) EDCs taken at ($\pi$, 0) using 29 and 50 eV excitation energies at which the contribution of the bonding states to the lineshape is negligible, thus unmasking the 'intrinsic' lineshape of the antibonding component. (d)-(l) Energy distribution maps taken straddling the (0,0)-($\pi$,0) alongs cuts parallel to ($\pi$, $\pi$) -- ($\pi$, 0) -- ($\pi$, -$\pi$).}
\end{figure}

Given such detailed photon energy dependent data, an attractive possibility now is to try and identify conditions for which the emission from one of the bilayer split bands is negligibly weak, thus offering access to the intrinsic line shape of the other band (plus background). As can be seen from Fig.~\ref{Spectra}, the matrix element from the bonding band has a local minimum for both doping levels at h$\nu$=29 eV and 50 eV. Keeping in mind that the bonding band lies much deeper in energy (260 meV) than the antibonding band, this effect is further multiplied by the strong broadening induced by the (frequency-dependent) self-energy. Thus, for 29 and 50 eV photon energies, the contribution of the bonding band to the ($\pi$,0) spectral line shape is vanishingly small. We replot the relevant spectra from the UD77K sample in Figs.~\ref{EDCs&EDMs}(a) and \ref{EDCs&EDMs}(b) and compare them with the h$\nu$ =50 eV OD76K spectrum in Fig.~\ref{EDCs&EDMs}(c). The difference between the spectra from the two doping regimes is subtle yet very clear: both EDCs from the underdoped system possess plateau or dip-like features, whereas the overdoped sample evidently exhibits a single component lineshape. We
stress that as the effects of the bilayer splitting have been effectively excluded for these conditions, the lineshape seen in Figs.~\ref{EDCs&EDMs}(a) and (b) is an intrinsic property of the spectral function of the antibonding CuO band. Single spectral function peak-dip-hump lineshapes are generally discussed in terms of
coupling between the electrons and a collective mode 
\cite{Norman,Abanov,Schrieffer,Scalapino}, whereby anomalies are expected in the electronic spectrum at energies where the probability for boson-mediated scattering of the electrons is maximal. In general, the mode energy can be read off
from the energetic separation between the peak and 'dip' (or plateau) in the ARPES lineshape \cite{Campuzano}. 
In this case the mode energy is between 38-40 meV and thus, bearing in mind the \textbf{k}-space location involved ($\pi$,0), one naturally begins to suspect the sharp resonance observed in INS as the role of the mediator of the scattering 
\cite{Norman}.

Returning to Fig.~\ref{EDCs&EDMs}(a-c), it would, of course, be tempting to conclude that the EDCs imply a strong doping dependence of the mode or of the coupling strength. However, as mentioned before, energetic locations of the bonding and antibonding bands change with doping. This brings with it the consequence that the antibonding band is simply too close to the Fermi level to be strongly influenced \cite{Gromko2}, implying in turn that the mode itself is sharply localised in energy. In order to be able to analyse the feedback effects as a function of doping we include into the consideration the bonding band. We show in Fig.~\ref{EDCs&EDMs} (d)-(l) energy distribution maps (EDMs) taken along ($\pi$, $\pi$) -- ($\pi$, 0) -- ($\pi$,-$\pi$) cuts in the Brillouin zone. It is convenient to refer to Fig.~\ref{Spectra}(b,c) when trying to identify the spectral features on the presented EDMs.
As discussed above (see Fig.~\ref{Spectra}(b,c)), for h$\nu$=50-55 eV (right-hand column in Fig.~\ref{EDCs&EDMs}) the data reflect predominantly the behavior of the antibonding band. For the other photon energies, the relative contribution from the bonding band can be much larger, which is particularly the case for h$\nu$=38 eV photons (center column of Fig.~\ref{EDCs&EDMs}). 

The collection of EDMs shown in Fig.~\ref{EDCs&EDMs} is an important and completely new set of ARPES data as they cover the ($\pi$,0) region of both the overdoped \textit{and} underdoped regimes for photon energies which differingly select the two bilayer-split bands. First, we mention that for the overdoped regime, the bilayer split bands are clearly visible in panels (d) and (e) giving rise to the 'large scale' ($\pi$,0) PDH. On going to the underdoped crystals, a qualitatively different picture emerges. The data from the superconducting state (middle row) appear to look very puzzling, with hardly any sign of the individual bilayer split bands, but rather a weakly dispersing, sharp feature located at $\sim$ 20-30 meV followed by an interval between 60 and 70 meV in which the spectral weight is strongly suppressed. Data collected above $T_c$ (bottom row) substantially clarify the situation: the picture is now remarkably similar to that of the overdoped case, with the two bilayer-split components being clearly seen to vary in relative intensity as the excitation energy changes, with the bonding band decreasing in strength on going from panel (k) to (j) to (l).

So now we return to the question: what happens to the electronic bands in the underdoped sample below $T_c$? Closer inspection of the low temperature EDMs reveals that considerable depletion of the spectral weight occurs for both the bonding and antibonding bands. What is also different from the overdoped case is that the two bands merge into one sharp and dispersionless feature above the energy of the 'dip' and are no longer distinguishable. We attribute such a difference in behaviour to the larger value of the gap and apparent stronger renormalization of the bonding band in the underdoped regime, both of which hamper the visual resolution of the bilayer splitting in this region of \textbf{k}-space. Changes in the dispersion of the bonding band have been observed before in overdoped samples \cite{Gromko1,kink}, but this is the first time the wholesale 'wipe-out' of spectral weight at energies some 30-40 meV below the peak feature near ($\pi$,0) has been shown in underdoped HTSC. Essentially this dramatic difference in the spectra corresponding to the bonding band argues for a strong doping dependence of the feeback effects caused by the bosonic mode and thus for the anomalous enhancement of the coupling strength upon underdoping.

\begin{figure}[b!]
\includegraphics[width=8.47cm]{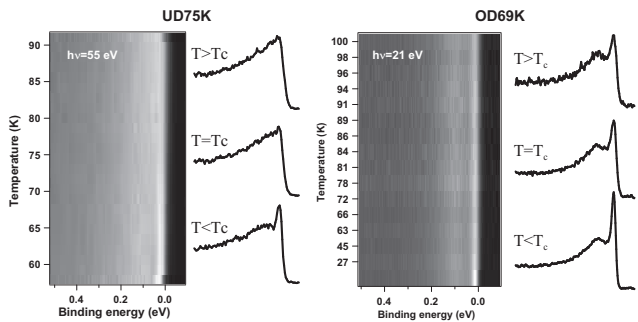}%
\caption{\label{T_dep} Temperature dependence of the peak-dip-hump structure in under- and overdoped cases.}
\end{figure}

The next step, then, is to determine whether the observed anomaly is linked to being in the superconducting state. To this end, we consider two cases in detail. Firstly, we show in the left panel of Fig.~\ref{T_dep} the temperature dependence of the ($\pi$,0) spectrum from an underdoped sample recorded using the photon energy at which intrinsic feature is seen. The dip (plateau) disappears approximately at $T_c$, which confirms its intimate relation with superconductivity. The right panel of the same figure shows the temperature dependence of the 'classic' PDH from the overdoped sample recorded using traditional h$\nu$ =21 eV. Upon overdoping, the PDH lineshape persists well above the $T_c$, which is fully consistent with both other experiments \cite{Gromko2,Yusof} and our interpretative framework in which the overdoped PDH is caused by a superposition of the bonding and antibonding bands \cite{KordPRL}.

Taking the ARPES data presented here in its entirety, we can now make a detailed inventary of the properties of the bosonic mode causing feedback effects in the electronic states. The 'fingerprints' of the mode are: its energy is about 38-40 meV; it only causes strong self-energy effects in the superconducting state; the mode coupling is maximal around ($\pi$,0) in momentum space and, finally, its influence is strongly doping dependent, being greatly enhanced in the underdoped regime. 
Considering the profile of the bosonic mode summarised in the preceeding, it is clear that it has to be identified with the sharp magnetic resonance mode observed in inelastic neutron scattering experiments, thus re-establishing the earlier arguments \cite{Norman,Abanov,Chubukov} and rebutting recent doubts \cite{Kee} in this regard.

In conclusion, we have presented a high resolution ARPES study 
of the interplay between bilayer splitting and intrinsic self energy effects
near ($\pi$,0) in underdoped bilayer cuprate superconductors. The self
energy effects are consistent with the interaction of electronic excitations
with a sharp bosonic mode.
By utilising a wide range of excitation energies, we are able to efficiently decouple the complicating effects of the bilayer splitting, thus enabling the identification of the key characteristics of the mode to which the electronic
states most intimitely involved with high T$_c$ superconductivity couple. The boson mode - which makes itself felt via a wipe-out of spectral weight giving rise to an intrinsic peak-dip-hump EDC lineshape - couples in significantly only below T$_c$ and does so much more strongly in the underdoped than in the overdoped regime. Furthermore, its location in energy (ca 38-40 meV) and k-space (at and near to [$\pi$,0]), taken together with the doping and temperature dependence unambiguously identify the boson as the magnetic resonance mode seen in inelastic neutron scattering.

We acknowledge stimulating discussions with Ilya Eremin, Dirk Manske and technical support from R. H\"ubel. HB is grateful to the Fonds National Suisse de la Recherche Scientifique and MSG to FOM for support.


\begin{thebibliography}{}
\bibitem{Rossat} J. Rossat-Mignod {\it et al.}, Physica C {\bf 185-189}, 86 (1991).
\bibitem{Tranquada} J. M. Tranquada {\it et al.}, Phys. Rev. B {\bf 46}, 5561 (1992).
\bibitem{Fong} H. F. Fong {\it et al.}, Nature {\bf 398}, 588 (1999).
\bibitem{Dai} P. Dai {\it et al.}, Nature {\bf 406}, 965 (2000).
\bibitem{Kee} H.-Y. Kee {\it et al.}, Phys. Rev. Lett. {\bf 88}, 257002 (2002).
\bibitem{Abanov} Ar. Abanov {\it et al.}, preprint cond-mat/0112126;
Ar. Abanov and A. Chubukov, Phys. Rev. Lett. {\bf 83},1652 (1999).
\bibitem{Norman} 
M. R. Norman and H. Ding, Phys. Rev. B {\bf 57}, R11089 (1998);
M. Eschrig and M. R. Norman, Phys. Rev. Lett. {\bf 85}, 3261 (2000), and
cond-mat/0202083.
\bibitem{Dessau}  D. S. Dessau \textit{et al.}, Phys. Rev. Lett. \textbf{66}, 2160 (1991).
\bibitem{BKJ} P. V. Bogdanov {\it et al.}, Phys. Rev. Lett. {\bf 85}, 2581 (2000); A. Kaminski {\it et al., ibid} {\bf 86}, 1070 (2001); P. D. Johnson {\it et al., ibid} {\bf 87}, 177007 (2001).
\bibitem{Zasad} J. F. Zasadzinski {\it et al.}, Phys. Rev. Lett. {\bf 87}, 067005 (2001).
\bibitem{Basov} D. N. Basov {\it et al.}, Phys. Rev. Lett. {\bf 77}, 4090 (1996).
\bibitem{Chubukov} A. V. Chubukov {\it et al.}, cond-mat/0201140.
\bibitem{Carbotte} J. P. Carbotte {\it et al.}, Nature {\bf 401}, 354 (1999), J. Orenstein ibid. {\bf 401}, 333 (1999).
\bibitem{KordPRL} A. A. Kordyuk \textit{et al.}, Phys. Rev. Lett. {\bf 89}, 077003 (2002).
\bibitem{Gromko1} A. D. Gromko {\it et al.}, cond-mat/0202329.
\bibitem{Gromko2} A. D. Gromko {\it et al.}, cond-mat/0205385.
\bibitem{Eschrig_Norman} M. Eschrig and M. R. Norman, cond-mat/0206544.
\bibitem{Kord_gap} A. A. Kordyuk {\it et al.}, cond-mat/0208418.
\bibitem{Footnote_PbDoping} The Pb doping suppresses the incommensurate modulation of the BiO layers, thus removing the diffraction replicas which otherwise contaminate the ($\pi$,0) region of \textbf{k}-space in ARPES.
\bibitem{gap} The only difference in the fitting procedure we applied here in comparison with the one, described in Ref.~\onlinecite{KordPRL} is that to account for the substantially large gap the binding energy scale was shifted by the value of the gap ($\sim$ 20meV).
\bibitem{Schrieffer} S. Engelsberg and J. R. Schrieffer, Phys. Rev. {\bf 131}, 993 (1963).
\bibitem{Scalapino} D. J. Scalapino in "Superconductivity", edited by R. D. Parks (Marcel Decker, New York, 1969).
\bibitem{Campuzano} J. C. Campuzano {\it et al.}, Phys. Rev. Lett. {\bf 83}, 3709 (1999).
\bibitem{kink} We note that although the doping level of our overdoped samples is lower (additionally defined as the area enclosed by their Fermi surfaces in the normal state) than those studied in Ref.~\onlinecite{Gromko1} the change in the dispersion upon entering the superconducting state is much weaker. We consider two possible factors which can account for such a discrepancy: our spectra are taken at higher temperatures and/or on Pb-doped single crystals where the superstructure replicas seen in Ref.~\onlinecite{Gromko1} are absent.
\bibitem{Yusof} Z. M. Yusof {\it et al.}, Phys. Rev. Lett. {\bf 88}, 167006 (2002).

\end{thebibliography}
\end{document}